\documentclass[twocolumn]{aa}
\usepackage{epsfig}
\usepackage{graphicx}

\begin{document}

% - new commands
\newcommand{\hi}{\ion{H}{i}~}
\newcommand{\hii}{\ion{H}{ii}~}

\title{High resolution observations of a starburst at z=0.223: resolved CO(1--0)  
structure\thanks{Based on observations carried out with the IRAM Plateau de Bure Interferometer.
IRAM is supported by INSU/CNRS (France), MPG (Germany) and IGN (Spain)}}

\author{F. Combes \inst{1}
\and
S. Garc\'{\i}a-Burillo \inst{2}
\and
J. Braine \inst{3}
\and
E. Schinnerer \inst{4}
\and
F. Walter \inst{4}
\and
L. Colina  \inst{5}
\and
M. Gerin \inst{6}
           }
\offprints{F. Combes}
\institute{Observatoire de Paris, LERMA (CNRS:UMR8112), 61 Av. de l'Observatoire, F-75014, Paris, France
\email{francoise.combes@obspm.fr}
 \and
Observatorio Astron\'omico Nacional (OAN)-Observatorio de Madrid,
Alfonso XII, 3, 28014-Madrid, Spain 
 \and
Observatoire de Bordeaux, Universit\'e Bordeaux~I, BP 89, 33270 Floirac, France
 \and
Max-Planck-Institut f\"ur Astronomie (MPIA), K\"onigstuhl 17, 69117 Heidelberg, Germany
 \and
IEM, Consejo Superior de Investigaciones Cientificas (CSIC), Serrano 121, 28006 Madrid, Spain
 \and
Radioastronomie ENS, 24 Rue Lhomond, 75005 Paris, France
              }

   \date{Received XXX 2006/ Accepted YYY 2006}

   \titlerunning{High resolution CO map of a starburst at z=0.223}
   \authorrunning{F. Combes et al.}

   \abstract{We present the results of CO(1--0) emission mapping with
the IRAM interferometer, at $\sim$ 1 \arcsec\,
resolution, of the z=0.223 ultra-luminous starburst IRAS
11582+3020. This galaxy was selected from an IRAM-30m survey of
30 galaxies at moderate redshift (z $\sim$ 0.2-0.6) to explore
galaxy evolution and in particular the star formation efficiency,
in the redshift range filling the gap between local and very
high-z objects. The CO emission is kinematically resolved, and
about 50\% of the total emission found in the 27\arcsec (97~kpc)
single dish beam is not recovered by the interferometer.
This indicates that some extended emission may be present on large scales
(typically 7-15\arcsec).  The FIR-to-CO luminosity ratio follows the trend
between local and high-z ultra-luminous starbursts.

}
\maketitle

%---------------------------------------------------------------

\section{Introduction}

The star formation history (SFH) of the Universe has been extensively
studied in recent years, through high spatial resolution observations of
star-forming galaxies as a function of look-back time 
(e.g. Madau et al 1998). The main striking feature
is a steady increase of the star formation rate between z=0 and z=1,
by at least a factor 10 or even more (Blain et al 1999a).

Starbursts were more frequent in the recent past, and star-forming
regions were also more dust enshrouded -- the evolution of galaxies
appears to be much faster in the infrared than in the optical/UV.
This strong evolution of infrared bright star-forming galaxies is not
really understood in models of galaxy evolution. It has been
established observationally that the number of galaxy mergers,
triggering starbursts, increases with redshift as a high power-law,
like $(1+z)^3$ (e.g. Le F\`evre et al 2000), and this is consistent with
numerical simulations and semi-analytical models (e.g. Balland et al 2003).
However to reproduce the strong SF evolution, the contribution
and efficiency of
mergers to the star-formation must also vary considerably with
redshift, with a peak at $z \sim 1$, to agree with the observations
(cf. Blain et al. 1999b, Combes 1999).

To better understand the physics responsible for this evolution, it is
paramount to measure the content and distribution of the fuel for star
formation, the molecular gas, as a function of redshift. Although a
few tens of ultra-luminous infrared galaxies, often amplified by a
gravitational lens, have been mapped in the CO line at very high
redshift (e.g. Omont et al 2003, 
Walter et al 2004, Tacconi et al 2006), not much is
known about star-forming galaxies at moderate distances, in the range 
$0.2 < z < 0.6$ (only 2 objects have been studied in this range,
Solomon et al 1997).  Therefore, we have undertaken a CO survey of 30
IR-luminous galaxies in this redshift range to check whether the
derived molecular gas content and star formation efficiency (SFE or
SFR/M(H$_2$)) are evolving at this faster rate as well.

For this study, we first obtained global information on the molecular
content and on the SFE using single dish data\footnote{An article
  presenting the results of our galaxy survey will be presented
  elsewhere, Combes et al.\ in prep.}. After a successful detection
in our single dish survey, we obtained high-resolution Plateau de Bure
Interferometer (PdBI) observations to map the molecular gas
distribution. Resolving the gas distribution and kinematics is
crucial to attribute the star formation activity to a global merger or a
less violent process. The first high-resolution CO observations
 of the survey have been
obtained towards IRAS 11582+3020 which is the object of this Letter.

\section{The sample and the source IRAS 11582+3020}

\begin{figure}[ht!]
\centering
\includegraphics[angle=0,width=8cm]{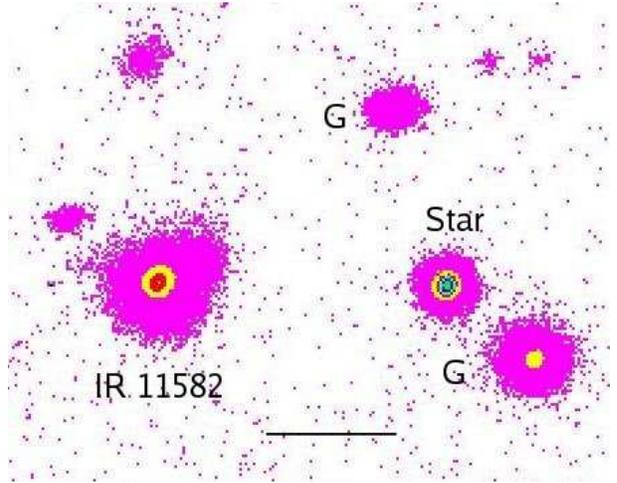}
\caption{Red image of IRAS 11582+3020 from Kim et al (2002).  The two
  objects marked "G" are galaxies in the same group (a fore-ground
  star is also indicated). The length of the horizontal bar at the
  bottom is 10\arcsec = 36~kpc at z=0.223.  }
\label{g4-red}
\end{figure}

The ULIRG sample of Solomon et al (1997) contains 37 objects, however
only 2 have $z > 0.2$. To fill the gap between low and high-z ($z>2$)
studies, we have initiated the first systematic survey of $ 0.2 < z <
0.6 $ sources, selecting the 30 most luminous galaxies detected at 
60 micron (IRAS or ISO) with known spectroscopic redshifts. 
IRAS 11582+3020 is one of the 12 well detected galaxies. It is an
ultra-luminous galaxy with L(IR)= 5.4 10$^{12}$ L$_\odot$ (with an IR
flux computed from IR= 1.8 x (13.48 f$_{12}$ + 5.16 f$_{25}$ +2.58
f$_{60}$+f$_{100}$) 10$^{-14}$ Wm$^{-2}$; Sanders \& Mirabel 1996).
Based on the ion lines in its optical spectrum it was classified as a
LINER by Kim et al (1998). The source is not detected in 2cm radio
continuum by the VLA, with an upper limit of 0.8 mJy (Nagar et al
2003), but it is a 20cm radio-source, with a flux of 3 mJy (Becker et
al 1995). The FIR-to-radio ratio 
$q$=log([F$_{FIR}$/(3.75 10$^{12}$ Hz)]/[f$_\nu$(1.4 GHz)])=2.7,
typical for ULIRGs (Sanders \& Mirabel 1996). 
Rupke et al (2005) find evidence for a superwind outflow in
this galaxy of about 15 $\rm M_{\odot}yr^{-1}$, while its SFR is
estimated at 740 $\rm M_{\odot}yr^{-1}$ from the infrared luminosity,
SFR= $\alpha$ L$_{IR}$ /(5.8 10$^9$L$_{\odot}$) (e.g. Kennicutt 1998),
using a correction factor of $\alpha$=0.8 for AGN contribution,
as adopted by Rupke et al (2005). The
red image of Kim et al. (2002) reveals some extended diffuse emission,
while two galaxies of the same group (according to the spectroscopy by
Veilleux et al 2002) are within 90~kpc in projection
(Fig~\ref{g4-red}). Veilleux et al. (2002) classify this system as a
post-merger, i.e. the tidal tails have been so diluted that they become
hardly visible, while the center is still perturbed with a prominent
knot of star formation.

In this article, we adopt a standard flat cosmological model,
with $\Lambda$ = 0.7, and a Hubble constant of 70\,km\,s$^{-1}$\,Mpc$^{-1}$.
IRAS 11582+3020 at z=0.223
is then at an angular distance of 743 Mpc,
and 1\arcsec = 3.6~kpc. The luminosity distance is 1111 Mpc.

\begin{figure}[ht!]
\centering
\includegraphics[angle=-90,width=8cm]{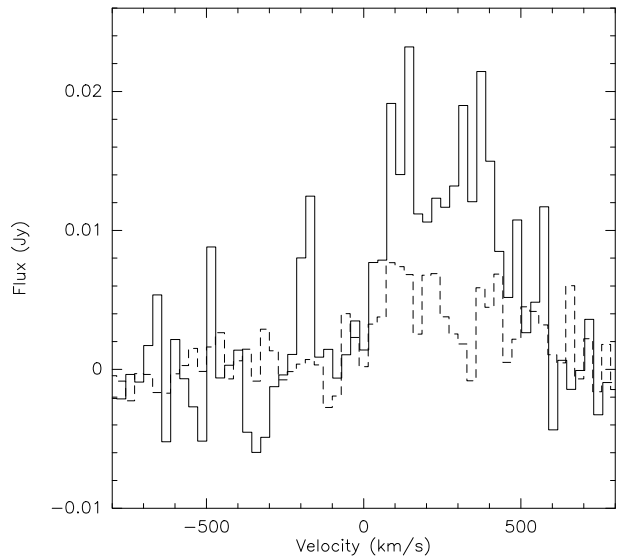}
\caption{The IRAM 30m CO(1--0) spectrum (solid line), compared
with the integrated spectrum from the interferometer (dash
line). The velocity is relative to z=0.223.}
\label{g4-30m}
\end{figure}

\begin{figure*}[ht!]
\centering
\includegraphics[angle=-90,width=8cm]{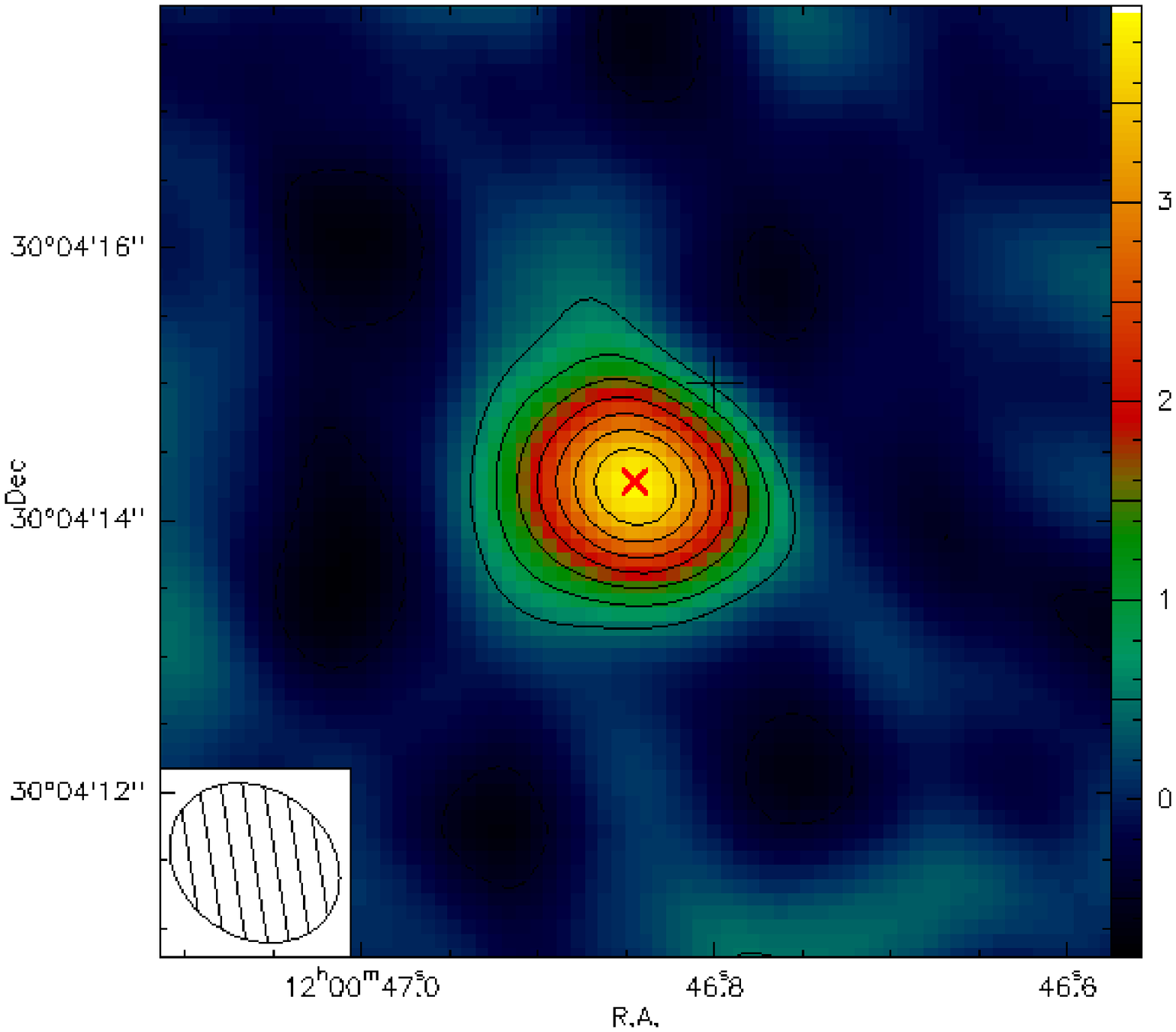}
\hspace{0.25cm}
\includegraphics[angle=-90,width=8cm]{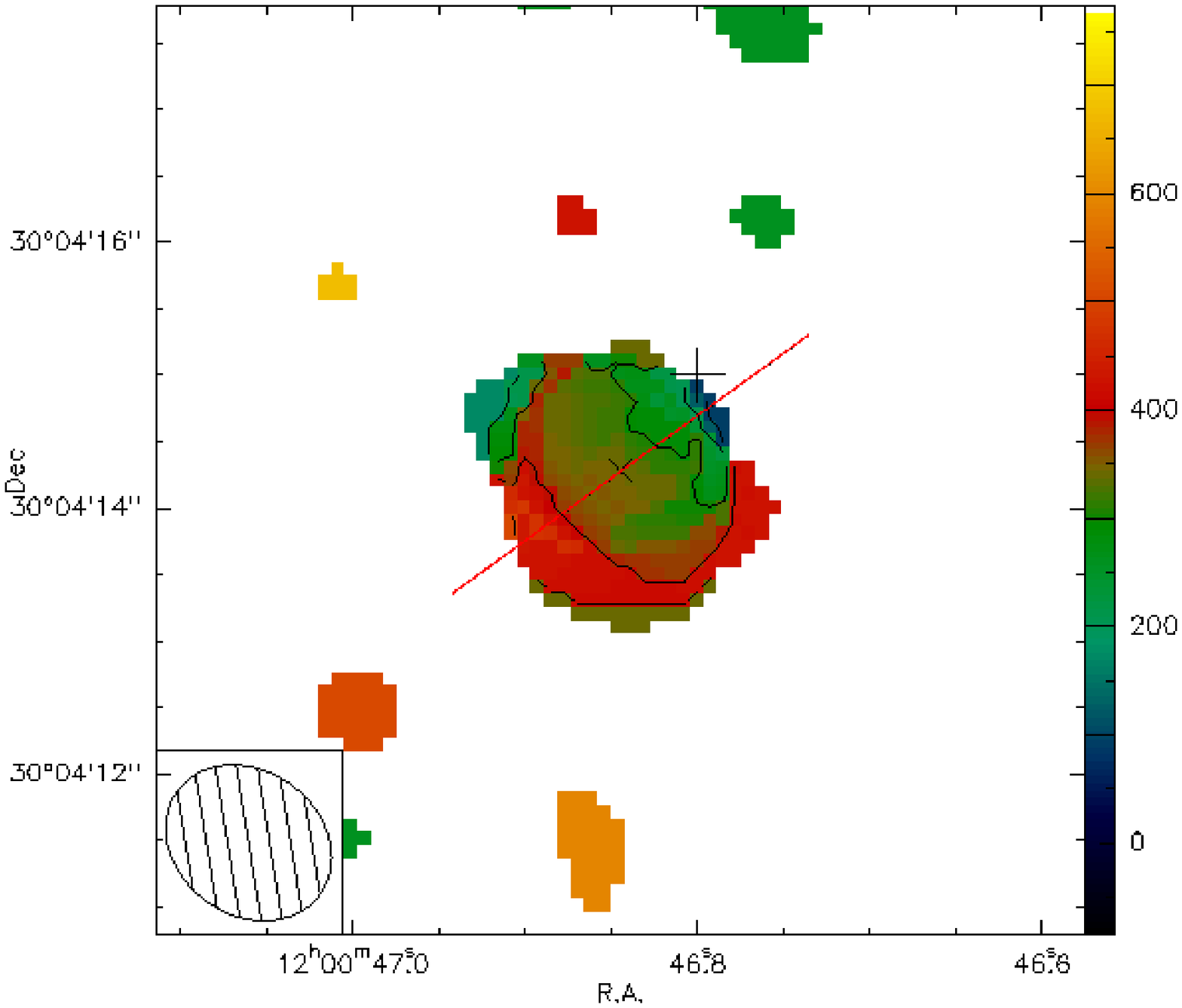}
\caption{{\bf a)} CO(1--0) integrated intensity contours in the inner
  $\sim$7$\arcsec$ (25.2~kpc) of IRAS~11582+3020. Contours are -0.5,
  0.5 to 4Jy\,km\,s$^{-1}$~beam$^{-1}$ in steps of
  0.5Jy\,km\,s$^{-1}$~beam$^{-1}$
  (1-$\sigma$=0.2Jy\,km\,s$^{-1}$~beam$^{-1}$). The black cross marks
  the position of the phase tracking
  center($\alpha_{J2000}$=$12^h00^m46.8^s$ and
  $\delta_{J2000}$=$30^{\circ}04'15.0''$). The red cross marks the
  position of the CO peak intensity and CO-based dynamical center
  ($\alpha_{J2000}$=$12^h00^m46.85^s$ and $\delta_{J2000}$=$30^{\circ}
  04'14.3''$). {\bf b)} The mean velocity field derived from the 1--0
  data using a 3$\sigma$-clipping in each channel map, in line
  contours, spanning the range (100~km~s$^{-1}$, 500~km~s$^{-1}$) in
  steps of 100~km~s$^{-1}$. The velocity scale is relative to z=0.223
  with a CO-derived systemic velocity of v$_{sys}$=300~km~s$^{-1}$,
  i.e., z$_{CO}$=0.224 (redhifted by 180 km\,s$^{-1}$ from the
optical line which is at z=0.2234). The red line shows the position of the
  apparent major axis of the galaxy. Beam-sizes are represented by
  hatched ellipses.}
\label{g4-flux}
\end{figure*}

\section{Observations}

First, we performed IRAM 30m observations in May 2005 (see
Fig.~\ref{g4-30m}). We used 2 SIS receivers to observe the 2
polarizations simultaneously at 94.253 GHz, the redshifted frequency
of the CO(1--0) line. At this frequency, the telescope half-power beam
width is 27$''$. The main-beam efficiency is $\eta$$_{\rm mb}=T_{\rm
  A}^*/T_{\rm mb}$=0.77, and $S/T_{\rm A}^*$ = 6.1 Jy/K. The typical
system temperature was 120 K (on the $T_{\rm A}^*$ scale). Wobbler
switching mode was used, with reference positions offset by 2$'$ in
azimuth. 
%The pointing was regularly checked on continuum sources and
%the accuracy was 3$''$ rms. 
Two 1 MHz filter banks provided
a total bandwidth of 512 MHz, or 1600 km\,s$^{-1}$, 
with a velocity resolution of 3.2 km\,s$^{-1}$.

IRAS 11582+3020 was subsequently observed with the 6 antennae of the 
PdBI in January and February 2006 in
the new A and B configurations. The 3mm receiver was tuned to 
94.253 GHz and the 1mm to 240 GHz for the dust continuum (as no
other CO line falls in the tuning bands). The
dual-band SIS receivers yielded SSB receiver
temperatures around 40\,K and 50\,K at the two observed frequencies.
The system temperatures were 120\,K for CO(1--0) and 400\,K at 240
GHz.  Four correlator units covered a total bandwidth of 580\,MHz at
each frequency, providing a nominal
frequency resolution of 1.25\,MHz ($4\,{\rm km\,s^{-1}}$ for the
CO(1--0) line), however we smoothed the signal to 28.6 km\,s$^{-1}$ channels. 
%Visibilities were obtained with twenty one-minute integrations on the
%source framed by short ($\sim2$\,min) phase and amplitude calibrations
%on the nearby quasars 0923+392, J0418+380 and 1156+295. The data were
%phase calibrated in the antenna-based mode. 
On average, the residual
atmospheric phase jitter was less than $20^\circ$ at 3mm, consistent
with a seeing disk of 0.34\arcsec--0.46\arcsec\ size and with a $\sim
5$\% loss of efficiency. 
%The fluxes of the primary calibrators were
%determined from IRAM measurements and taken as an input to derive the
%absolute flux density scales for our visibilities, estimated to be
%accurate to 10\%.

Reduction using the GILDAS software provided data
cubes with 512$ \times $512 spatial pixels (0.14"/pixel) and
64 velocity channels of 28.6 km\,s$^{-1}$ width. The cubes were cleaned
with the Clark (1980) method and restored by a 1.3\arcsec$\times$
1.0\arcsec\ Gaussian beam (with PA=54$^{\circ}$) at 94.253 GHz and
0.43\arcsec$\times$ 0.25\arcsec\ (with PA=44$^{\circ}$) at 240GHz.
The rms noise levels in the cleaned maps (at 28.6 km\,s$^{-1}$ velocity
resolution) are 0.8\,mJy\,beam$^{-1}$ for the CO(1--0) line. No continuum
emission was detected at 3mm (possible AGN) or at 1mm  (possible
dust emission), down to rms noise levels of 0.1 mJy\,beam$^{-1}$ and
0.5 mJy\,beam$^{-1}$ in a 580\,MHz bandwidth at 94 GHz and 240 GHz,
respectively. Given the IRAS 100$\mu$m flux of 1.5 Jy, the upper limit 
at 1.2mm is compatible with a typical starburst SED.

\section{Results}

As revealed in Fig~\ref{g4-30m}, the total PdBI flux is $\sim50$\% of
what is found with the single dish. When taking the intrinsic flux
uncertainties into account this may indicate that some extended
emission is not recovered by our interferometer observations. With a
beam of 27\arcsec\ at the 30m, we measured a CO(1--0) flux toward the
center of I(CO)= 1.3 K.km\,s$^{-1}$ (in the T$_A^*$ scale). With a conversion
factor of 6.1 Jy/K, the integrated flux is then S(CO)=7.8 Jy.km\,s$^{-1}$.
With the PdBI, we measure 3.7 Jy.km\,s$^{-1}$ in the FOV= 54\arcsec\, of the
CO(1--0) map. This can be explained by the long PdBI baselines which
are not sensitive to CO emission extending over 10\arcsec\, and detect
only the more clumpy distribution. The optical structure does reveal
the existence of such an extended component, and also hints at a weak
diffuse tidal tail in this perturbed system, classified as a
post-merger (Veilleux et al 2002).

The CO luminosity for a high-z source is given by $$L_{CO} =
23.5 I_{CO} \Omega_B {{D_L^2}\over {(1+z)^3}} \hskip6pt \rm{K\hskip3pt
  km \hskip3pt s^{-1}\hskip3pt pc^2}$$
where $\Omega_B$ is the area of
the main beam in square arcseconds and $D_L$=1111 is the luminosity
distance in Mpc.  We compute H$_2$ masses by $M_{H_2} = \alpha L_{CO}$
M$_\odot$, with $\alpha=0.9$ for ULIRGs, instead of 4.6 for normal
galaxies (i.e. Solomon et al. 1997). The molecular masses are
listed in Table~\ref{table_H2}.

\begin{figure}[t]
\centering
\includegraphics[angle=0,width=8cm]{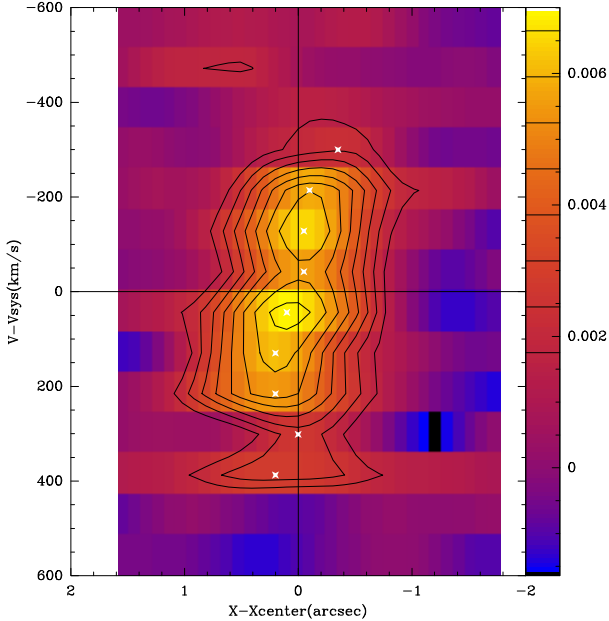}
\caption{We show the CO(1--0) position-velocity diagram along
  the major axis of IRAS~11582+3020 (line contours range: 2.5$\sigma$,
  to 10.5$\sigma$ in steps of 1$\sigma$; 1$\sigma$=0.7~mJy~beam$^{-1}$
  in channels of 86~km~s$^{-1}$). Velocities have been re-scaled to
  z$_{CO}$=0.224 and angular offsets are relative to the dynamical
  center.  The white crosses mark the velocities corresponding to the
  peak brightness temperatures as a function of position.}
\label{major-g4}
\end{figure}

The integrated CO(1--0) map is plotted in Fig~\ref{g4-flux}, together
with the isovelocity curves. The source is resolved at least in the
direction of the beam minor axis, where the deconvoled size of the CO
emitting region is of the order of 0.8\arcsec\, in diameter ($\sim$
3~kpc). The signal-to-noise ratio is around 10 in most of the channels,
implying a precision on the position of the peaks of
$\sim$0.1\arcsec\,. Therefore it is possible to detect the shift of
the barycenter in each channel map, with a kinematic major axis
aligned at PA  $\sim$135$^\circ$ (Fig~\ref{g4-flux}). The velocity
gradient is also clearly seen in the position-velocity diagram 
taken along this position angle (Fig~\ref{major-g4}). Since $\sim$135$^\circ$
is also the position angle of the extended optical isophotes (Fig~\ref{g4-red}),
all available data are compatible with a post-merger relaxed system.

The CO integrated spectrum seems to show a double-horn profile, indicative of
a rotating disk. 
%Another possibility is the presence of two galaxies,
%however this is unlikely, given the apparent evolved state of
%IRAS\,11582+3020 as a post-merger. 
Since the contours of the red
image are suggesting an inclination of i=50$^\circ$ with the plane of
the sky for a disk geometry, we can estimate a dynamical mass,
 assuming that the width of the CO profile is twice the 
projected rotational velocity V$_{rot}$ sin i = 250 km\,s$^{-1}$. With V$_{rot}$
= 326 km\,s$^{-1}$, inside a radius of 1.5~kpc, i.e. the extent of the central
molecular disk, the indicative dynamical mass is M$_{dyn}$ = 3.4
10$^{10}$ M$_{\odot}$.  With the low conversion ratio proposed for the
ULIRGs, the gas mass in the central molecular disk (M(H$_2$) = 6
10$^{9}$ M$_\odot$) is a small fraction ($\sim$ 20\%) of the dynamical
mass inside 1.5~kpc.

\begin{table}[t]
\centering
\begin{tabular}{lccc}
\hline
\hline
Instrument & $\Delta$V & Area & M(H$_2$) \\
            & km\,s$^{-1}$ & Jy.km\,s$^{-1}$ &   M$_\odot$ \\
\hline \rule{0pt}{2.6ex}
30m  & 476$\pm$33& 7.8$\pm$0.5  &  1.2$\pm$0.1 10$^{10}$  \\
\rule{0pt}{2.6ex}
PdBI  & 550$\pm$52 & 3.7$\pm$0.3  &  6$\pm$0.4 10$^{9}$  \\
\hline
\end{tabular}
\caption{\small  CO results from IRAM-30m and PdBI, and deduced
H$_2$ mass ($\Delta$V is the FWHM).
}
\label{table_H2}
\end{table}

\section{Discussion and conclusions}

The $\sim$ 1\arcsec\, resolution CO observations of the
ultra-luminous starburst IRAS 11582+3020 at $z=0.223$ provides
interesting insight into its molecular gas content and morphology. The
CO emission is resolved, spatially and kinematically: we see evidence
for a rotating disk. In addition to a central molecular disk of
$\sim$1.5~kpc in radius, there is tentative evidence for additional
extended emission over 10-15\arcsec\, as about 50\% of the 30m flux
are not recovered by the interferometer. Such an extended component
could correspond to the perturbed optical morphology of the evolved
merger. This source size contrasts with the 10 nearby ULIRGs, mapped by
Downes \& Solomon (1998) where the molecular gas is more concentrated,
with typical radii of 300 to 800 pc. The derived total amount of
molecular gas in IRAS 11582+3020 is comparable to
the average mass in local ULIRGs of $\sim$ 5 10$^9$ M$_\odot$ and its
M$_{gas}$/M$_{dyn}$ ratio is also comparable to the value of $\sim$
1/6 found by Downes \& Solomon (1998).

The physical properties of IRAS 11582+3020 appear to be intermediate
between that of local ULIRGs and high-z submillimeter galaxies
(SMG) mapped by Tacconi et al (2006). The latter have also a typical
radius of 2~kpc, and their FIR-to-CO luminosity ratio is higher than
the local ULIRGs ratio, due to their higher FIR luminosities. 
There might be a trend, also followed by IRAS 11582+3020, for high-z
galaxies to have a larger L(FIR)/M(H$_2$) ratio (Riechers et al 2006),
however this must be confirmed with larger samples.
 The high-z SMG of Tacconi et al (2006) are only
mapped in the higher J rotation lines of CO, which may explain the
complete absence of extended emission, even in the case of mergers.
From their study of the three CO-brightest z$\ge$4 QSOs, Riechers et
al (2006) limit a potential extended CO emission to $<$30\% of the total.
Two of their quasars are gravitationally lensed (with magnification
factors of $\sim$3--7), i.e. differential magnification may slightly
increase this limit.

%%%%%%%%%%%%%%%%%%%%%%%% acknowledgments
\begin{acknowledgements}
The authors gratefully acknowledge P. Salom\'e for his help
in the interferometric data reduction and D.-C. Kim for having provided
the red and K' images of IRAS 11582+3020.
We have made use of the NASA/IPAC Extragalactic Database
(NED).
\end{acknowledgements}
%%%%%%%%%%%%%%%%%%%%%%%%%%%%%%%%%%%%%


\begin{thebibliography}{}
\bibitem{}Balland C., Devriendt J., Silk J.: 2003, MNRAS 343, 107
\bibitem{}Becker R.H., White R.L., Helfand D.J.: 1995 ApJ 450, 559
\bibitem{}Blain A.W., Smail I., Ivison R.J., Kneib J-P.: 1999a, MNRAS 302, 632
\bibitem{}Blain A.W., Jameson A., Smail I. et al. : 1999b, MNRAS 309, 715
%\bibitem{}Carilli C.L., Lewis G.F., Djorgovski S.G. et al.: 2003, Science 300, 773
\bibitem{}Combes F.: 1999, Ap\&SS 269, 405 (astro-ph/9909016)
\bibitem{}Downes D., Solomon P.: 1998, ApJ 507, 615
\bibitem{}Kennicutt R.C.: 1998, ApJ 498, 541
\bibitem{}Kim D.C., Veilleux S., Sanders D.B.: 1998, ApJ 508, 627
\bibitem{}Kim D.C., Veilleux S., Sanders D.B.: 2002, ApJS 143, 277
\bibitem{}Le F\`evre O., Abraham R., Lilly S.J. et al.: 2002, MNRAS 311, 565
%\bibitem{}Lilly S.J., Le F\`evre O., Hammer F., Crampton D.: 1996, ApJ 460, L1
%\bibitem{}Madau P., Ferguson H.C., Dickinson M.E. et al.: 1996, MNRAS 283, 1388
\bibitem{}Madau P., Pozzetti L., Dickinson M.E.: 1998, ApJ 498, 106
\bibitem{}Nagar, N.M., Wilson, A.S., Falcke, H. et al: 2003 A\&A 409, 115
\bibitem{}Omont A., Cox P., Beelen A., Bertoldi F., Carilli C.L.: 2003, AGN from
     Central Engine to Host Galaxy, PASP, 290, 583
\bibitem{}Riechers, D.A., Walter, F., Carilli, C. et al.: 2006, AJ, 650, 604
\bibitem{}Rupke D.S., Veilleux S., Sanders D.B.: 2005 ApJS 160, 115
\bibitem{}Sanders D.S., Mirabel F., 1996, ARAA, 34, 749
\bibitem{}Solomon P., Downes D., Radford S., Barrett J.: 1997, ApJ 478, 144
\bibitem{}Tacconi L.J., Neri R., Chapman S.C. et al.: 2006, ApJ 640, 228
\bibitem{}Veilleux S., Kim D-C., Sanders D.B.: 2002, ApJS 143, 315
\bibitem{}Walter F., Carilli C., Bertoldi F. et al.: 2004, ApJL 615, L17
\end{thebibliography}
\end{document}